\input{psfig}
\input{epsf}
\documentstyle[12pt]{article}

\def\dofig#1{\vskip.2in\centerline{\epsfbox{#1}}}

\def\simge{\mathrel{%
   \rlap{\raise 0.511ex \hbox{$>$}}{\lower 0.511ex \hbox{$\sim$}}}}
\def\simle{\mathrel{
   \rlap{\raise 0.511ex \hbox{$<$}}{\lower 0.511ex \hbox{$\sim$}}}}
 
\def\slashchar#1{\setbox0=\hbox{$#1$}           
   \dimen0=\wd0                                 
   \setbox1=\hbox{/} \dimen1=\wd1               
   \ifdim\dimen0>\dimen1                        
      \rlap{\hbox to \dimen0{\hfil/\hfil}}      
      #1                                        
   \else                                        
      \rlap{\hbox to \dimen1{\hfil$#1$\hfil}}   
      /                                         
   \fi}                                         %
\def\ts{\thinspace}

\def\ra{\rightarrow}

\def\ol{\bar}

\def\be{\begin{equation}} 
\def\ee{\end{equation}} 
\def\bea{\begin{eqnarray}}
\def\eea{\end{eqnarray}}
\def\ba{\begin{array}}
\def\ea{\end{array}}
\def\CA{{\cal A}}
\def\CB{{\cal B}}
\def\CC{{\cal C}}

\def\CM{{\cal M}}

\def\CO{{\cal O}}

\def\etmiss{\slashchar{E}_T}

\def\ecm{\sqrt{s}}
\def\shat{\hat s}

\def\rshat{\sqrt{\shat}}

\def\atro{\alpha_{\tro}}

\def\Ntc{N_{TC}}

\def\thw{\theta_W}
\def\kslash{\raise.15ex\hbox{/}\kern-.57em k}
\def\tro{\rho_{T}}

\def\tropm{\rho_{T}^\pm}

\def\troz{\rho_{T}^0}
\def\tpi{\pi_T}
\def\tpipm{\pi_T^\pm}
\def\tpimp{\pi_T^\mp}
\def\tpip{\pi_T^+}
\def\tpim{\pi_T^-}
\def\tpiz{\pi_T^0}

\def\ptjj{p_T(jj)}
\def\dfjj{\Delta\phi(jj)}
\def\wjj{Wjj}

\def\jet{\rm jet}
\def\jets{\rm jets}

\def\gev{{\rm GeV}}
\def\tev{{\rm TeV}}

\def\pb{{\rm pb}}
\def\ipb{{\rm pb}^{-1}}

\def\ifb{{\rm fb}^{-1}}
\def\half{{\textstyle{ { 1\over { 2 } }}}}
\def\third{{\textstyle{ { 1\over { 3 } }}}}
\def\twothirds{{\textstyle{ { 2\over { 3 } }}}}

\begin{document}
\title{
\vskip -15mm
\begin{flushright}
\vskip -15mm
{\small FERMILAB-PUB-97/116-T\\
BUHEP-97-13\\
hep-ph/9704455\\}
\vskip 5mm
\end{flushright}
{\Large{\bf \hskip 0.38truein
Finding Low-Scale Technicolor \hfil\break at Hadron Colliders}}\\
}
\author{
{\small Estia Eichten$^{1}$\thanks{eichten@fnal.gov},}
{\small Kenneth Lane$^{2}$\thanks{lane@buphyc.bu.edu}, and}
{\small John Womersley$^{1}$\thanks{womersley@fnal.gov}}\\
{\small {$^{1}$}Fermilab, P.O.~Box 500, Batavia, IL 60510}\\
{\small {$^{2}$}Dept.of Physics, Boston University, 
590 Commonwealth Avenue, Boston, MA 02215}\\
}
\maketitle
\begin{abstract}
In multiscale and topcolor-assisted models of walking technicolor,
relatively light spin-one technihadrons $\tro$ and $\omega_T$ exist and are
expected to decay as $\tro \ra W\tpi$, $Z\tpi$ and $\omega_T \ra \gamma
\tpi$. For $M_{\tro} \simeq 200\,\gev$ and $M_{\tpi} \simeq 100\,\gev$, these
processes have cross sections in the picobarn range in $\ol p p$ colisions
at the Tevatron and about 10~times larger at the Large Hadron Collider. We
demonstrate their detectability with simulations appropriate to Run~II
conditions at the Tevatron.
\end{abstract}


\newpage

\section{Introduction}

Light, color-singlet technipions, $\tpipm$ and $\tpiz$, are expected to
occur in models of multiscale technicolor \cite{multi} and
topcolor-assisted 
technicolor~\cite{topcondref,topcref,tctwohill,tctwoklee}. These
technipions will be resonantly produced via technivector meson dominance at
substantial rates at the Tevatron and the Large Hadron
Collider~\cite{tpitev,snow}. The technivector mesons in question are an
isotriplet of color-singlet $\tro$ and the isoscalar partner $\omega_T$.
Because techni-isospin is likely to be a good approximate symmetry, $\tro$
and $\omega_T$ should have equal masses as do the various technipions. The
enhancement of technipion masses due to walking technicolor~\cite{wtc}
suggest that the channels $\tro \ra \tpi\tpi$ and $\omega_T \ra
\tpi\tpi\tpi$ are closed. Thus, the decay modes $\tro \ra W_L \tpi$ and
$Z_L \tpi$, where $W_L$, $Z_L$ are longitudinal weak bosons, and $\omega_T
\ra \gamma \tpi$ may dominate~\cite{multi}. We assume in this paper that
$\tpiz$ decays into $b \ol b$. These heavy flavors, plus an isolated
lepton or photon, provide the main signatures for these processes.

We present simulations of $\ol p p \ra \tropm \ra W^\pm_L \tpiz$
and $\omega_T \ra \gamma \tpiz$ for the Tevatron collider with $\sqrt{s} =
2\,\tev$ and an integrated luminosity of $1\,\ifb$. For $M_{\tro} \simeq
200\,\gev$ and $M_{\tpi} \simeq 100\,\gev$, cross sections at the Tevatron are
expected to be several picobarns. The narrowness of the $\tro$ and
$\omega_T$ resonances suggests topological cuts that enhance the
signal-to-background ratio. For the cross sections we assume in this paper,
the signals stand out well above the background once a single $b$-tag is
also required. Although we do not simulate these processes for the LHC, cross
sections there are an order of magnitude larger than at the Tevatron, so
detection of the light technihadrons should be easy. We focus first on
$\tro$ production and decay and take up the $\omega_T$ later.

\section{$\tropm \ra W^\pm \tpiz$}

In Ref.~\cite{tpitev}, we assumed that there is just one light isotriplet
and isoscalar of color-singlet technihadrons and used a simple model of
technirho production and decay to determine the rates of the processes
\begin{equation}
\label{eq:singlet}
\begin{array}{llll}
q \ol q' & \ra W^\pm & \ra \tropm &\ra \ts\ts W_L^\pm Z_L^0;  \quad W_L^\pm
\tpiz, \ts\ts \tpipm Z_L^0;  \quad \tpipm \tpiz \\ \\
q \ol q & \ra \gamma, Z^0 & \ra \troz &\ra \ts\ts W_L^+ W_L^-; \quad
W_L^\pm \tpimp; \quad \tpip \tpim \;
\end{array}
\end{equation}
and their dependence on $M_{\tro}$. We found that the most important
processes were those with positive $Q = M_{\tro} -$ (sum of final-state
masses) {\it and} the fewest number of longitudinal weak bosons. For
$M_{\tro} \simle 250\,\gev$ and $M_{\tpi} \simeq 100\,\gev$, the dominant
processes have cross sections of 1--10~pb at the Tevatron and 10--100~pb
at the LHC. The $Q$-values of the processes that dominate production
typically are less than 30--40~GeV.

Technipion production at hadron colliders is based on a
vector-meson-dominance model. The subprocess cross sections for a pair of
technipions $\pi_A \pi_B$, including longitudinal weak bosons, are
\begin{equation}
\label{eq:trhocross}
{d\hat\sigma(q_i \ol q_j \ra \tro^{\pm,0} \ra \pi_A\pi_B)
\over{d(\cos\theta)}} =  {\pi \alpha^2 p_{AB}^3 \over
{3 \shat^{\textstyle{{5\over{2}}}} }}
\ts {M^4_{\tro} \ts \sin^2\theta \over
{(\shat - M_{\tro}^2)^2 + \shat \Gamma_{\tro}^2(\shat)}} \ts
A_{ij}^{\pm,0}(\shat) \ts \CC^2_{AB} \ts,
\end{equation}
where $\alpha$ is the fine-structure constant, $\shat$ is the
subprocess center-of-mass energy, $p_{AB}$ is the technipion momentum, and
$\theta$ is the $\pi_A$ production angle in the subprocess
c.m.~frame. Ignoring Kobayashi-Maskawa mixing angles, the factors
$A_{ij}^{\pm,0} = \delta_{ij} A_i^{\pm,0}$ are (for $\shat \gg M_W^2$,
$M_Z^2$)
\be\label{eq:afactors}
\ba{lll}
&A_i^\pm(\shat) &= {1 \over {4 \sin^4\thw}} \biggl({\shat \over {\shat -
M_W^2}}\biggr)^2
\ts, \\ \\
&A_i^0(\shat) &= \vert \CA_{iL}(\shat) \vert^2
+ \vert \CA_{iR}(\shat) \vert^2 \ts; 
\\ \\
&\CA_{iL}(\shat) &= Q_i + {2 \cos 2\thw \over {\sin^2 2\thw}} \ts
(T_{3i} - Q_i \sin^2\thw) \biggl({\shat \over {\shat - M_Z^2}}\biggr)\ts,
\\
&\CA_{iR}(\shat) &= Q_i - {2 Q_i \cos 2\thw \sin^2\thw \over{\sin^2
2\thw}} \ts \biggl({\shat \over {\shat - M_Z^2}}\biggr) \ts. \;
\ea
\ee
Here, $Q_i = \twothirds, \ts -\third$ and $T_{3i} = \half, -\half$ are the
electric charge and third component of weak isospin for (left-handed)
quarks~$u_i$ and $d_i$, respectively.

The energy-dependent width $\Gamma_{\tro}(\shat)$ in
Eq.~(\ref{eq:trhocross}) is
given by the sum of the $\tro$ partial widths to technipions $\pi_A \pi_B$
and to fermion pairs $\ol f_i f_j$. The former is
\begin{equation}
\label{eq:trhowidth}
\Gamma(\tro \ra \pi_A \pi_B) = {2 \atro \CC^2_{AB}\over{3}} \ts
{\ts\ts p_{AB}^3\over {\shat}} \ts,
\end{equation}
where the coupling $\atro = 2.91 (3/\Ntc)$ is naively scaled from QCD.
In calculations, we take the number of technicolors to be $\Ntc =4$. The
parameter $\CC^2_{AB}$ depends on a mixing angle $\chi$ and is given
by~\cite{tpitev,multi}
\be\label{eq:ccab}
\ba{ll}
\CC^2_{AB} &= \left\{\ba{ll} \sin^4\chi  & {\rm for} \ts\ts\ts\ts W_L^+ W_L^-
\ts\ts\ts\ts {\rm or} \ts\ts\ts\ts  W_L^\pm Z_L^0 \\
\sin^2\chi \cos^2\chi & {\rm for} \ts\ts\ts\ts W_L^+ \tpim, W_L^- \tpip
\ts\ts\ts\ts  {\rm or} \ts\ts\ts\ts W_L^\pm \tpiz, Z_L^0 \tpipm \\
\cos^4\chi & {\rm for} \ts\ts\ts\ts \tpip\tpim  \ts\ts\ts\ts {\rm or} \ts\ts
\ts\ts \tpipm\tpiz 
\ea \right.
\ea
\ee
Note that, for a given technirho, $\sum_{AB} \CC^2_{AB} = 1$.
We take $\sin\chi = \third$ in our calculations.
The $\tro$ are {\it very} narrow, $\Gamma(\tro\ra\pi_A\pi_B) \simle
1\,\gev$ for $\rshat=M_{\tro} \simeq 210\,\gev$. The decay rates of the 
$\tro$ to fermion-antifermion states are even smaller and are given by
\begin{equation}
\label{eq:trhoqq}
\Gamma(\tro^{\pm,0} \ra \ol f_i f_j) = {C_f \ts \alpha^2 M^4_{\tro}\over
{3 \atro\shat^2}} \ts {p_i \ts (2 \shat^2-\shat(m^2_i+m^2_j)  - (m^2_i -
m^2_j)^2) \over{2 \shat^2}}   \ts A_{ij}^{\pm,0}(\shat) \ts,
\end{equation}
where $C_f = 3$ (1) for color-triplet (singlet) final-state fermions;
$p_i$ is the momentum and $m_i$ the mass of fermion $f_i$.

We focus on technirho decay modes with the best signal-to-background
ratios, namely, $\tro \ra W_L \tpi$ or $Z_L \tpi$. For definiteness, we
assume $M_{\tro} = 210\,\gev$ and $M_{\tpi} = 110\,\gev$. Technipion
couplings to fermions are expected to be proportional to
mass. We adhere to that expectation in this paper and assume that
technipions decay as
\be\label{eq:singdecay}
\ba{ll}
\tpiz &\ra b \ol b \\
\tpip &\ra c \ol b \ts\ts\ts {\rm or} \ts\ts\ts c \ol s, \ts\ts
\tau^+ \nu_\tau \ts.
\ea
\ee
Thus, heavy-quark jet tagging is an important aid to technipion
searches.\footnote{We note that some topcolor-assisted technicolor
models~\cite{tctwoklee} have the feature that certain technifermions, and
their bound-state technipions, couple mainly to the lighter fermions of the
first two generations. The flavor-blind kinematical cuts we discuss below
will be essential for this possibility.}

We have used {\sc Pythia}~6.1~\cite{pythia} to
generate $\ol p p \ra W^\pm \ra \tropm \ra W^\pm \tpiz$ with $\tpiz \ra \ol
b b$ at the Tevatron Collider with $\ecm = 2\,\tev$. The cross section for
this process is $5.3\,\pb$. We also used {\sc Pythia} to generate the
$W^\pm \ts\jet\ts\jet$ background. 
Jets were found using the clustering code provided in {\sc Pythia}
with a cell size
of $\Delta\eta\times\Delta\phi = 0.1\times 0.1$, a cone radius $R=0.7$ and
a minimum jet $E_T$ of 5 GeV. Cell energies were smeared using a 
calorimeter resolution
of $0.5\sqrt{E{\rm (GeV)}}$. Missing transverse energy $\etmiss$
was then calculated as:
\begin{equation}
\etmiss = - \left\vert \sum_{\jets,\ell^\pm} {\bf E_T}\right\vert
\end{equation}
Selected
events were required to have an isolated electron or muon, large missing
energy, and two or more jets. The selection criteria were: 
\begin{itemize}
\item Lepton: $E_T(\ell) > 25\,\gev$; pseudorapidity $|\eta| < 1.1$.
\item Missing energy: $\etmiss > 25\,\gev$.
\item Transverse mass: $50\,\gev < \CM_T(\ell \etmiss) < 100\,\gev$.
\item Two or more jets with $E_T > 20\,\gev$
and $|\eta| < 2.0$, separated from the lepton by at least $\Delta R = 0.7$.
\end{itemize}

\begin{figure}
\epsfxsize=15cm
\dofig{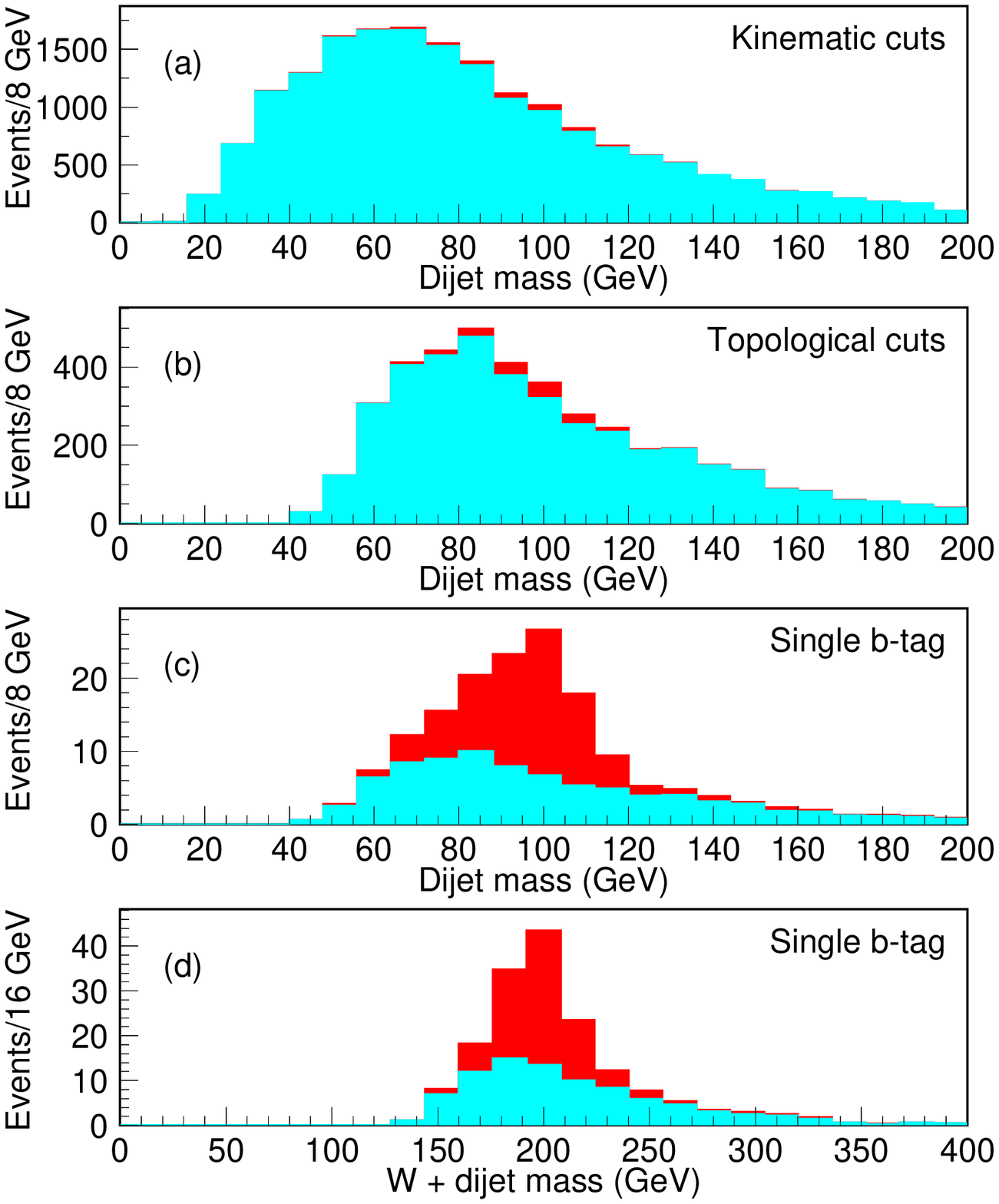}
\caption[]
{Invariant mass distributions for
$\tro$ signal (black) and $Wjj$ background (grey);
vertical scale is events per bin in 1~fb$^{-1}$ of
integrated luminosity.
Dijet mass distributions (a) with 
kinematic selections only, (b) with the addition 
of topological selections,
and (c) with the addition of single $b$-tagging; 
(d) $W+$dijet invariant mass distribution for the same
sample as (c).
\label{tech_rho1}}
\end{figure}

Requiring that the lepton and jets be central in pseudorapidity exploits
the fact that the signal events will tend
to be produced with larger center-of-mass
scattering angles than the background.  
Figure~1(a) shows the invariant mass distribution of the two highest-$E_T$
jets for the signal (black) and background (grey) events passing these
criteria for a luminosity of $1\,\ifb$, half that expected in Tevatron
Run~II.

The peculiar kinematics of $\tro \ra W_L \tpi$ and $Z_L \tpi$ suggest other
cuts that can discriminate signals from the $W/Z + \jets$
backgrounds.\footnote{The following discussion applies to both the Tevatron
and  the LHC but, because of the smaller boost rapidities of the $\tro$ at
the Tevatron, signal events will be more central there. Cutting harder on
rapidity for LHC events will improve signal-to-background; the higher cross
section and luminosity at the LHC leave plenty of events.} The small
$Q$-values for $\tro$ decays causes the $\tpi$ and $W_L$ (or $Z_L$) to have
low transverse momenta, $p_T < p_T^{\rm max} = \sqrt{M^4_{\tro} -
2M^2_{\tro}(M^2_{\tpi} + M^2_W) + (M^2_{\tpi} - M^2_W)^2} \simeq 45\,\gev$
for our reference masses. Not only is the $p_T$ of the dijet system
limited, but the jets are emitted with an opening azimuthal angle $\dfjj
\simge 140^\circ$. These expectations were borne out by simulated
distributions in these variables. Cutting on the maximum and minimum
$\ptjj$ and the minimum $\dfjj$ help suppress the $\wjj$ background to
$\tro \ra W_L \tpi$.

Consequently, we have taken the selected events in Fig.~1(a) and applied
the additional topological cuts $\dfjj > 125^\circ$ and $20 < \ptjj <
50\,\gev.$ These cuts reject 78\% of the $Wjj$ background while retaining
64\% of the signal. The results are shown in Fig.~1(b). For the signal
cross section of $5.3\,\pb$, the signal-to-background at $100\,\gev$ is
improved from 0.04 to 0.11 by these cuts. A signal rate in excess of
$15\,\pb$ would produce a visible excess at this level.

The additional effect of tagging one $b$-jet per event is shown in
Fig.~1(c).  We have assumed a 50\% efficiency for tagging $b$'s, a 1\%
probability to mistag light quarks and gluons, and a 17\% probability to
mistag charm as a $b$.   This final selection leaves a clear dijet
resonance signal above the background at just below the mass of the $\tpi$.
For $80<m_{jj} < 120\,\gev$, there are 65~signal events over a background
of~35. The mass distribution for the signal is almost gaussian,  with a
peak at $97\,\gev$ and $\sigma \simeq 12.7\,\gev$. This width and a tail on
the low side are due mainly to the effects of final-state gluon radiation,
fragmentation, calorimeter resolution and neutrinos from $b$-decay.

Figure~1(d) shows the invariant mass distributions for the $Wjj$ system
after topological cuts and $b$-tagging have been imposed.
Here the $W$ four-momentum was reconstructed from the lepton and $\etmiss$,
taking the lower-rapidity solution in each case.  Again, a clear peak
is visible at just below the mass of the $\tro$. We point out
that, especially after making the topological cuts, the dijet mass and
the $Wjj$ mass are highly correlated; a peak in one distribution is
almost bound to correspond to a peak in the other.  Thus, the existence of
structure in both distributions does not add to the statistical significance
of any observation.

We conclude that if $\tro$ and $\tpi$ exist in the mass range we consider,
they can be found without difficulty in Run~II of the Tevatron. In fact,
since there are $\sim 65$ signal events in Fig.~1(c) for $\sigma(\tro^\pm
\ra W^\pm \tpiz) \simeq 5\,\pb$, one might see hints of a signal with one
tenth the luminosity. It is certainly worth looking in the presently
accumulated samples of $\sim100\,\ipb$ per experiment. The production cross
section at $1.8\,\tev$ is about 15\% lower than at $2.0\,\tev$. Other
channels may also add to the cross section: for our reference masses,
$\tro^0 \ra W^\pm \tpimp$ contributes an additional $2.3\,\pb$, though only
one $b$-jet is present in the $\tpipm$ decay.

\section{$\omega_T \ra \gamma \tpiz$}

We turn now to the signatures for $\omega_T$ production. The $\omega_T$ is
produced in hadron collisions just as the $\tro^0$, via its
vector-meson-dominance coupling to $\gamma$ and $Z^0$. Its cross
section is proportional to $|Q_U + Q_D|^2$, where $Q_{U,D}$ are the
electric charges of the $\omega_T$'s constituent technifermions. For
$M_{\omega_T} \simeq M_{\tro}$, then, the $\omega_T$ and $\troz$ should be
produced at comparable rates, barring accidental cancellations. If the
$\tro \ra \tpi\tpi$ channels are nearly or fully closed, then $\omega_T
\ra \tpi\tpi\tpi$ certainly is forbidden. If we can use decays of the
ordinary $\omega$ as a guide, $\omega_T \ra \gamma \tpiz$, $Z^0 \tpiz$ will
be much more important than $\omega_T \ra \tpi\tpi$~\cite{tpitev}. It is not
possible to estimate the relative importance of these two modes without an
an explicit model, but it seems plausible that $\gamma \tpiz$ will dominate
the phase-space-limited $Z^0 \tpiz$ channel. Therefore, we concentrate on
the $\omega_T \ra \gamma\tpiz \ra \gamma b \ol b$ mode in this paper. We
shall find, surprisingly enough, that the decay modes $\omega_T \ra \ol q
q$ and $\ell^+ \ell^-$ decay modes are comparable to $\gamma
\tpiz$.\footnote{We thank Torbjorn Sjostrand for emphasizing to us the
importance of the $\omega_T \ra \ell^+ \ell^-$ channel.}

As for the technirho, the subprocess cross section for $\omega_T \ra
\gamma\tpiz$ production is obtained from a simple vector-meson-dominance
model:
\begin{equation}
\label{eq:tomegacross}
{d\hat\sigma(q_i \ol q_i \ra \omega_T \ra \gamma\tpiz)
\over{d(\cos\theta)}} = {\pi \alpha^3 p^3 \over{12 \atro M^2_T \ts
\shat^{\textstyle{{3\over{2}}}} }} 
\ts {M^4_{\omega_T} \ts \ts (1+\cos^2\theta) \over
{(\shat - M_{\omega_T}^2)^2 + \shat \Gamma_{\omega_T}^2(\shat)}} \ts
B_i^0(\shat) \ts,
\end{equation}
where $p$ is the photon momentum. The factor $B_i$ is given by
\be\label{eq:bfactor}
\ba{ll}
B_i^0(\shat) &= \bigl\vert \CB_{iL}(\shat) \bigr\vert^2
+ \bigl\vert \CB_{iR}(\shat) \bigr\vert^2 \ts; \\ \\
\CB_{iL}(\shat) &= \biggl[Q_i - {4 \sin^2\thw \over {\sin^2 2\thw}} \ts
(T_{3i} - Q_i \sin^2\thw) \biggl({\shat \over {\shat -
M_Z^2}}\biggr)\biggl] \ts (Q_U + Q_D)
\ts,\\ \\
\CB_{iR}(\shat) &= \biggl[Q_i + {4 Q_i \sin^4\thw \over{\sin^2 2\thw}}
\ts \biggl({\shat \over {\shat - M_Z^2}}\biggr)\biggr] \ts (Q_U + Q_D)
\ts.
\ea
\ee
Neglecting $\omega_T \ra Z^0 \tpiz$, the energy-dependent width
$\Gamma_{\omega_T}(\shat)$ in Eq.~(\ref{eq:tomegacross})
is given by the sum of the partial widths
\be\label{eq:tomegadecay}
\ba{ll}
&\Gamma(\omega_T \ra \gamma \tpiz) = {\alpha p^3 \over {3 M_T^2}} \ts,
\\ \\
&\Gamma(\omega_T \ra \ol f_i f_i) = {C_f \ts \alpha^2 M^4_{\omega_T} \over
{3 \atro\shat^2}} \ts {p_i\ts (\shat - m^2_i) \over
{\shat}}\ts B^0_i(\shat) \ts.
\ea
\ee
The mass parameter $M_T$ in the $\omega_T \ra \gamma\tpiz$ rate is unknown
{\it a priori}; we take it to be $M_T = 100\,\gev$ in our calculations. We
also took the technifermion charges to be $Q_U ={4\over 3}$ and $Q_D = Q_U
- 1 = {1\over 3}$, and we used $M_{\omega_T} = M_{\tro} = 210\,\gev$  and
$M_{\tpi} = 110\,\gev$.

Note that the decay rate for the $\gamma\tpiz$ mode is $\CO(\alpha)$ and
the differential subprocess cross section is nominally of $\CO(\alpha^3)$.
Nevertheless, the integrated $p^\pm p \ra \omega_T \ra \gamma\tpiz$ rate
will be comparable to the rate for $p^\pm p \ra \tro\ra W_L\tpi$ and
$Z_L\tpi$ {\it if} the $\gamma\tpiz$ mode dominates the $\omega_T$ width.
Note also that the two rates in Eq.~(\ref{eq:tomegadecay}) are numerically
comparable when summed over fermion flavors. 

{\sc Pythia}~6.1 \cite{pythia} was used to generate $\ol p p \ra \gamma,
Z^0 \ra \omega_T \ra \gamma \tpiz$ with $\tpiz \ra \ol b b$ at $\ecm =
2\,\tev$. The cross section for this process is $2.6\,\pb$. The background
considered is $\gamma \ts\jet\ts\jet$. Selected events were required to
have an isolated photon and two or more jets. The selection criteria were:

\begin{itemize}
\item Photon: $E_T(\gamma) > 50\,\gev$; pseudorapidity $|\eta| < 1.1$.
\item Two or more jets with $E_T > 20\,\gev$
and $|\eta| < 2.0$, separated from the photon by at least $\Delta R = 0.7$.
\end{itemize}

\begin{figure}
\epsfxsize=15cm
\dofig{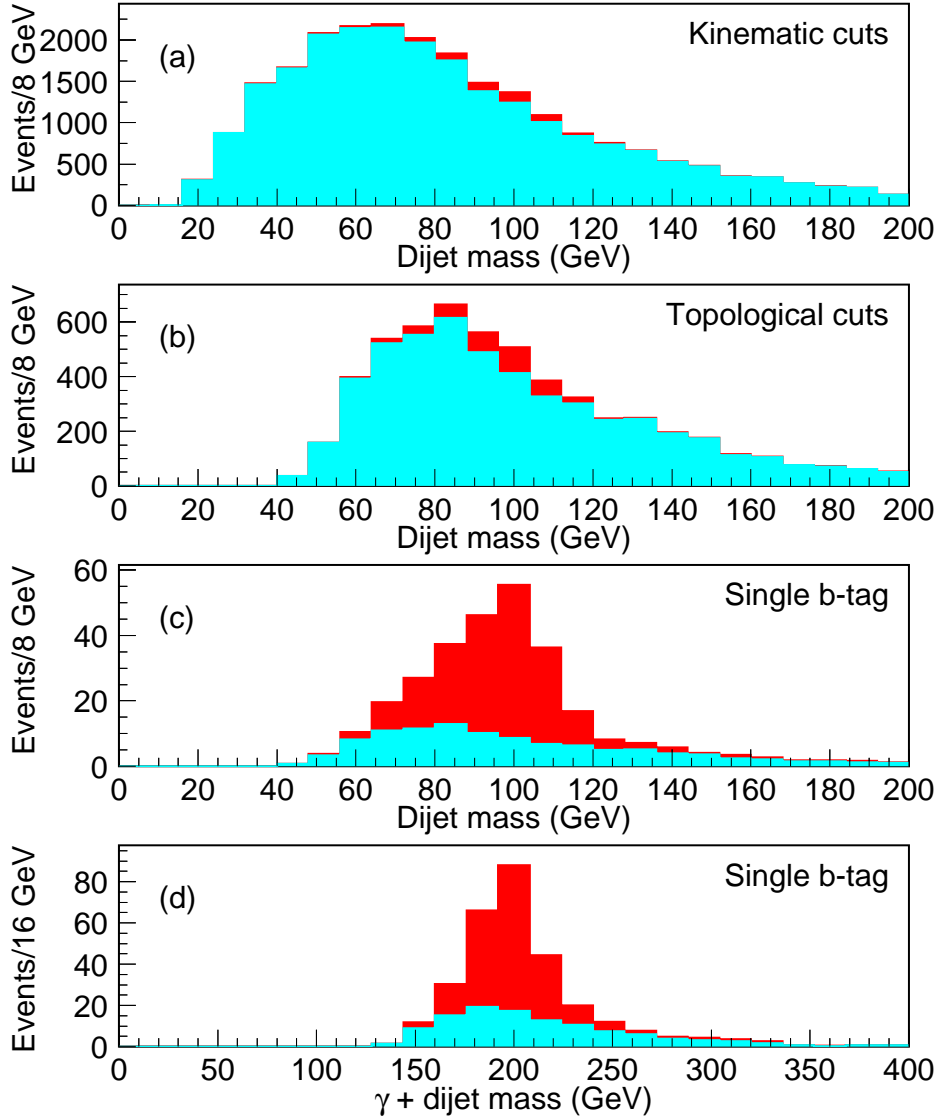}
\caption[]
{Invariant mass distributions for
$\omega_T$ signal (black) and $\gamma jj$ background (grey);
vertical scale is events per bin in 1~fb$^{-1}$ of
integrated luminosity.
Dijet mass distributions (a) with 
kinematic selections only, (b) with the addition 
of topological selections,
and (c) with the addition of single $b$-tagging; 
(d) $\gamma+$dijet invariant mass distribution for the same
sample as (c).
\label{tech_omega1}}
\end{figure}

Figure~2(a) shows the invariant mass distribution of the two highest-$E_T$
jets for the signal (black) and background (grey) events passing these
criteria for an integrated luminosity of $1\,\ifb$. As for the $\tro$,
backgrounds swamp the signal. Therefore, we have again investigated
topological selections which might enhance the signal over the backgrounds.
For $M_{\omega_T} \simeq 200\,\gev$ and $M_{\tpi} \simeq 100\,\gev$, signal
events have $p_T \simle 75\,\gev$ and dijet azimuthal angle $\dfjj \simge
105^\circ$.
We apply the additional cut $\dfjj > 90^\circ$ to the untagged events of
Fig.~2(a). However, no useful cut can made on $\ptjj$ since the signal and
background have very similar shapes.

The effect of these cuts is seen in Fig.~2(b). The $\gamma jj$ background
is reduced by 61\% while 75\% of the signal is retained. Tagging one
$b$-jet further improves the signal/background as shown in Fig.~2(c), and
a clear peak just below the $\tpi$ mass can be seen.  Figure~2(d) shows the
photon$+$dijet invariant mass after the topological cuts and $b$-tagging
are employed. Again, we found that this total invariant mass was not a
useful variable to cut on.

In conclusion, we have shown that the low-scale technicolor signatures
$\tro \ra W \tpi$ and $\omega_T \ra \gamma \tpi$ can be discovered easily
in Run~II of the Tevatron for production rates as low as a few picobarns.
Low rates require $b$-tagging of the $\tpi$. Topological cuts alone---which
may be the only handle for technipions decaying to charmed or lighter quark
jets---will be sufficient if cross sections exceed 10--15~pb. Signals with
rates this large should be apparent in the existing data. We urge that they
be looked for. Production rates are an order of magnitude higher at the LHC
than at the Tevatron. Thus, the LHC will be decisive in excluding low-scale
technicolor signatures of the type considered here.

We are greatly indebted to Torbjorn Sjostrand for including the $\tro$ and
$\omega_T$ processes in {\sc Pythia} and for many helpful and stimulating
comments. The research of EE and JW is supported by the Fermi National
Accelerator Laboratory, which is operated by Universities Research
Association, Inc., under Contract~No.~DE--AC02--76CHO3000. KL's research is
supported in part by the Department of Energy under
Grant~No.~DE--FG02--91ER40676. KL thanks Fermilab for its hospitality
during various stages of this work.

\end{document}